%% file: latticeUED-fin.tex
\documentclass[
preprint,
showpacs, preprintnumbers, amsmath, amssymb, aps, prd,floatfix,
nofootinbib,eqsecnum]{revtex4}
\usepackage{graphicx}
\usepackage{bm} 
\usepackage{bbm}

\include{comandaments}

\newcommand{\kk}{\ensuremath{B^0\rightleftharpoons\overline{B}^0}}

\newcommand{\Fig}[1]{Fig.~\ref{#1}}
\newcommand{\Eq}[1]{Eq.~\rf{#1}}
\newcommand{\Sec}[1]{Sec.~\ref{#1}}

\newcommand{\currentscalevalue}{0.6}
\newcommand{\comment}[1]{}

\newcommand{\ok}{
}

\begin{document}

\preprint{
\begin{tabular}{r}
FTUV--03--0630\\ 
IFIC/03--31 
\end{tabular}
}
\title{Bounds on models with one latticized extra dimension}

\author{J.F. Oliver}
\affiliation{Departament de F\'{\i}sica Te\`orica and IFIC, Universitat de
Val\`encia -CSIC\\
Dr. Moliner 50, E-46100 Burjassot (Val\`encia), Spain}
\author{J. Papavassiliou}
\affiliation{Departament de F\'{\i}sica Te\`orica and IFIC, Universitat de
Val\`encia -CSIC\\
Dr. Moliner 50, E-46100 Burjassot (Val\`encia), Spain}
\author{A. Santamaria}
\affiliation{Departament de F\'{\i}sica Te\`orica and IFIC, Universitat de
Val\`encia -CSIC\\
Dr. Moliner 50, E-46100 Burjassot (Val\`encia), Spain}

\date{\today}

\begin{abstract}
We study an extension of the standard model with
one latticized extra dimension accessible to all fields.
The model is characterized by the size of the extra dimension and the
number of sites, and contains a tower of massive particles.
At energies lower
than the mass of the new particles there are no tree-level effects. Therefore,
bounds on the scale of new physics can only be set from one-loop processes. 
We calculate several observables 
sensitive to loop-effects, such as the
$\rho$ parameter, $b\to s \gamma$, $Z\to b\overline b$, and the
$B^0\rightleftharpoons\overline{B}^0$ mixing, and use them to set
limits on the lightest new particles for different number of sites.
It turns out that the continuous result is rapidly reached when the
extra dimension is discretized in about 10 to 20 sites only. For small
number of sites  the bounds placed on the usual continuous
scenario can be reduced by roughly a factor of $10\%$--$25\%$, 
which means that the new particles can be as light 
as $320~\mbox{GeV}$.  Finally, we briefly
discuss an alternative model in which fermions do not have additional
modes.
\end{abstract}
\maketitle

\section{Introduction}
The interest on the possible existence of additional spatial dimensions 
\cite{Arkani-Hamed:1998rs, Antoniadis:1998ig, Arkani-Hamed:1998nn} 
has been renewed in the last years, when 
it was realized that many long standing
problems in particle physics and cosmology could be addressed from an entirely
different perspective. The mass spectrum of the fermions
\cite{Cheng:1999fw, Yoshioka:1999ds, Arkani-Hamed:1999za,
Rizzo:2001cy, Chang:2002ne, Biggio:2003kp}, novel neutrino oscillation
scenarios \cite{Dienes:1998sb, Arkani-Hamed:1998vp, Dvali:1999cn,
Ioannisian:1999cw, Barenboim:2001wy}, possible grand unification at
low scales \cite{Dienes:1998vg, Hebecker:2002vm, Oliver:2003cy}, as 
well as new ways to understand the family puzzle \cite{Dobrescu:2001ae},
are just some examples of the fields
were extra dimensions could be relevant.

One common prediction of extra-dimensional scenarios is the existence
of a tower of Kaluza-Klein (KK) modes on top of each degree of freedom
propagating in the bulk.
The
towers modify the low energy predictions, a fact that has been exploited
to set bounds on the size of the extra dimensions. Many precision
observables sensitive to these modifications have been used in the
literature for the case of continuous extra dimensions \cite{Masip:1999mk,
Papavassiliou:2000pq, Appelquist:2000nn, Agashe:2001ra, Agashe:2001xt, Muck:2001yv,
Buras:2002ej, Oliver:2002up, Buras:2003mk}.
In fact, the lowest KK states, if sufficiently light, could be 
produced in the next generation of accelerators. 
As far as this last possibility is concerned, the models with 
universal extra dimensions (UED)~\cite{Appelquist:2000nn}, i.e. 
with all the Standard Model (SM) fields propagating in the bulk,
seem to be particularly promising. 
In these models, 
due to the conservation of the KK-number (momentum-conservation in the
extra dimension) the only effects at low energy (i.e. below the
threshold of production of new particles) arise at one loop.
This allows for substantially lower bounds, compared to other models:
the masses of the new particles can be as low as $400$~GeV without
contradicting present experimental data.

Gauge theories in extra dimensions are not renormalizable and suffer from
the standard ambiguities when one tries to use them beyond their range of 
applicability~\cite{Oliver:2003cy}. 
This has motivated the search of theories that are better behaved 
at high energies, and reduce to extra-dimensional models at low energy
(or simulate sufficiently their spectrum and interactions). Among the possible
candidates we will concentrate on the so-called ``deconstructed extra
dimensions''\cite{Arkani-Hamed:2001ca} and 
``latticized extra dimensions''\cite{Hill:2000mu, Cheng:2001vd}. 
The former constitute UV completions of higher dimensional field theories, 
with gravity decoupled: at very high energies one 
starts
with  particularly constructed 
four-dimensional theories, which are renormalizable,
and in most cases even asymptotically free.  
Then an extra (latticized)  
dimension is generated dynamically at low energies, through the 
condensation of fermions, which transform appropriately   
under the various gauge groups~\cite{Arkani-Hamed:2001nc}. 
In the context of these theories a new way for
solving the hierarchy problem has been pointed out~\cite{Arkani-Hamed:2001nc,
Arkani-Hamed:2002qy}; the Higgs is understood as a pseudo-Goldstone
boson associated to a symmetry that has an extra-dimensional
analogue. The most economical form of these models, known under the name
of ``little Higgs'' models, are currently the object of an intense study, mainly 
in order to determine  to which extent the cancellation of quadratic divergences 
can be achieved without  fine-tunning \cite{Csaki:2002qg}.
On the other hand, the latticized  theories~\cite{Hill:2000mu, Cheng:2001vd} 
focus on a manifestly gauge-invariant effective 
Lagrangian description 
of the KK modes in 3+1 dimensions. 
Such a description is particularly useful when dealing with 
non-abelian gauge theories, 
because it evades complications 
with gauge-invariance, arising 
when hard momentum cutoffs are used 
in loop expansions, which  
is essentially what the usual truncation of the KK-tower 
amounts to.
The common feature of both types of theories, at least for our purposes,
is that they mimic to some extent an extra-dimensional behavior and share an
interpretation in terms of a discretized extra dimension. 

An important phenomenological difference between continuous
and latticized scenarios is the structure of the KK tower.
Specifically, 
whereas in the continuous scenarios the KK towers are
infinite, in the latticized versions they are finite, due to the
presence of a minimum physical distance, namely the distance between
sites in the extra dimensions\footnote{In latticized scenarios the
replicated degrees of freedom are not usually called KK modes,
nevertheless we will call them simply ``\emph{modes}'' since this
stresses the similarity between latticized and continuous cases.}. 
As commented above, in the case of 
universal extra dimensions the bounds on the masses of the lowest KK-particles
is rather low, a fact which offers the challenging
possibility of (pair)-producing them in upcoming experiments.
Therefore, it is important to study how this picture changes when the
extra dimensions are latticized. 
In this paper we perform an analysis, similar to that
of the continuous cases, to determine
the modifications to the bounds on the masses of new particles 
when one universal extra dimension is latticized. 

As the continuum theory, the latticized version of the universal extra 
dimensional model has no new tree-level effects at low energy, and bounds
can only be set from loop processes. We will try to place limits on the new
physics scale by studying several well measured quantities that in the 
standard model depend strongly on the top-quark mass\footnote{It is natural
to look for dependences on the top-quark mass because those are less
suppressed. In fact, if the new physics decouple, as it is the case,
the scale of new physics should appear in the denominator of physical 
observables. This scale must be compensated
by another mass and the largest mass available in the SM is the top-quark mass}: 
the $\rho$ parameter,
$b\to s \gamma$, $Z\to b\overline b$ and the
$B^0\rightleftharpoons\overline{B}^0$ mixing. 
We focus on the dominant pieces of the radiative corrections 
for a large top-quark mass. 

For comparison we discuss briefly some results for an alternative scenario
in which fermions do not propagate in the latticized extra dimension. In this
model the effects of new particles appear already at tree level. In
particular, they modify the Fermi constant which allow for much stronger
limits on the masses of new particles. Some one-loop processes also provide
interesting bounds but they are not competitive with the tree-level bounds.   

The paper is organized as follows: in \Sec{themodel} we describe a
full latticized version of the standard model focusing on the relevant
pieces for our calculations, the electroweak sector in general and top-quark
couplings in particular. In \Sec{Bounds} the contributions of the new
physics are computed for the different observables and the bounds one can
set on the mass of new physics particles are discussed. 
In \Sec{Conclusions} we summarize the main results of 
our study. Finally, in Appendix~A we collect some details of the
derivation of the spectrum of the model.

\section{The model}
\label{themodel}
In this section we will  
specify the field content of the model and its Lagrangian, and 
extract the mass spectrum 
and couplings necessary for computing the relevant 
observables. 

Following \cite{Hill:2000mu,Cheng:2001vd}, the Lagrangian is 
given by
\begin{equation}
\mathcal{L}=\mathcal{L}_G+\mathcal{L}_F+\mathcal{L}_H+\mathcal{L}_Y \, ,  
\end{equation}
where the various pieces are defined as follows.
The gauge piece, $\mathcal{L}_G$, associated to the gauge
group\footnote{We will not consider the contributions associated with
strong interactions.} $G=\Pi_{i=0}^{N-1} SU(2)_{i}\times U(1)_{iY}$, 
reads 
\begin{eqnarray}
\mathcal{L}_G&=&\sum_{i=0}^{N-1}-\frac{1}{4}F_{i\mu\nu}^{a}F^{i\mu\nu
a}- \frac{1}{4}F_{i\mu\nu}F^{i\mu\nu} \nonumber\\
\label{eq:kinterm}
  & +&\sum_{i=1}^{N-1}\mbox{Tr}\{(D_\mu\Phi_i)^\dagger (D^\mu \Phi_i) \}+ 
(D_\mu\phi_i)^\dagger (D^{\mu}\phi_i) - V(\Phi,\phi)\,,
\label{eq:potential_term}
\end{eqnarray}
where $F_{i\mu\nu}^a$ is the strength tensor associated with the gauge
field of the i-th $SU(2)_{i}$ and $F_{i\mu\nu}$ is the one for
$U(1)_{iY}$. $\Phi_i$ and $\phi_i$ are elementary scalars that will
acquire a VEV (common to every $i$), due to the potential term
$V(\Phi,\phi)$. 
Each of them become effectively nonlinear $\sigma$
model fields that can be parametrized as usual in terms of the scalar
fields $\pi_i$ and $\pi^a_i$
\begin{equation}
\phi_i=\frac{v_1}{\sqrt{2}}e^{i\pi_i/v_1}\, \qquad\qquad
\Phi_i=v_2 e^{i\pi^a_i \tau^a/2 v_2}\, , 
\end{equation}
where $v_1$ and $v_2$ are the VEVs of $\phi_i$ and $\Phi_i$ respectively and
$\tau^a$ are the usual Pauli matrices. In this paper we will
concentrate on the so called ``\emph{aliphatic model}''
\cite{Hill:2000mu},  
in which the $\Phi_i$ fields 
transform as $\left({2}_i,\bar{2}_{i-1}\right)$ under
the groups $SU(2)_{i}$ and $SU(2)_{i-1}$, as singlets for the
rest, and carry no $U_{iY}(1)$ charge, while the $\phi_i$
fields are singlets under all the $SU(2)$ groups and they are charged
only under $U(1)_{iY}$ and $U(1)_{(i-1)Y}$ 
(all hypercharges $Y_i$ will be eventually set to $Y_i=1/3$ ~\cite{Cheng:2001vd}). 
Thus, the covariant derivative assumes the form
\begin{equation}
D_\mu \Phi_i = \partial_\mu \Phi_i - i \mathcal{W}_{\mu,i} \Phi_i +
i \Phi_i \mathcal{W}_{\mu,i-1}\,,
\end{equation}
where $\mathcal{W}_{\mu,i}=\tilde{g} W_{\mu\;i}^a T^a_i$, $T^a_i$ are
the generators of the $SU_{i}(2)$ and $\tilde{g}$ is the
dimensionless gauge coupling constant that is assumed to be the same
for all the $SU(2)$ groups. The $U(1)$ covariant derivative for the $\phi_i$ can
be constructed similarly. 

The fermionic piece, $\mathcal{L}_F$, contains the following fields
(generational indices are suppressed)
\begin{equation}
\label{eq:fielddefinition}
Q_i=\left[\begin{array}{c}Q_{ui}\\
Q_{di}\end{array}\right]\,,
\qquad
U_i \,, 
\qquad 
D_i\,, 
\qquad 
i=0,\ldots,N-1 
\end{equation}
$Q_i$ transforms as a doublet under $SU(2)_i$ and as a singlet for the
rest of $SU(2)$ groups, and among the $U(1)$ fields it is only charged
under $U(1)_i$, with hypercharge $Y_Q=1/3$. 
$U_i$ and $D_i$ are only charged under $U(1)_i$,
with hypercharges $Y_U=4/3$ and $Y_D=-2/3$.
They are all vector-like fields with right- and left-handed chiral
components except for $i=0$. In this case $Q$
is left-handed and $U$ and $D$ are right-handed, which is equivalent
to imposing $Q_{0R}=0$, $U_{0L}=0$, and $D_{0L}=0$.
Then one sets $\mathcal{L}_F=\mathcal{L}_Q+\mathcal{L}_U+\mathcal{L}_D$, where
\begin{eqnarray}
\mathcal{L}_{Q} &=& 
\sum_{i=0}^{N-1}
\overline{Q}_{iL} i\Slash{D} Q_{iL} +\overline{Q}_{iR} i\Slash{D}Q_{iR}
- M_f \overline{Q}_{iL} \bigg(\frac{\sqrt{2} \Phi_{i+1}^{\dagger}
\phi_{i+1}^{\dagger}}{v_1 v_2} Q_{i+1R}-Q_{iR}\bigg)+\hc \,,\nonumber\\
\mathcal{L}_{U}&=& 
\sum_{i=0}^{N-1} \overline{U}_{iR} i\Slash{D} U_{iR} +
\overline{U}_{iL} i\Slash{D} U_{iL} +
M_f \overline{U}_{iR}
\bigg(\frac{\phi_{i+1}^{4\dagger}}{(v_1/\sqrt{2})^4}
U_{i+1L}-U_{iL}\bigg)+\hc\,,
\label{eq:fer5deri2a}
\end{eqnarray}
and $\mathcal{L}_D$ can be extracted from $\mathcal{L}_U$ making the next
substitutions, $U\to D$, $\phi_i\to \phi_i^{\dagger}$ and the exponent
should be replaced $4\to 2$. $\Slash{D}$ is
the usual covariant derivative associated with the gauge group
$G$, and $M_f$ is a generic mass that in principle could depend on $i$ but
usually it is set equal for all of them. 

The next piece in the Lagrangian, $\mathcal{L}_H$, is the one
associated with the Higgs doublet \cite{Cheng:2001vd}
\begin{equation}
\mathcal{L}_H = \sum_{i=0}^{N-1}(D_\mu H_i)^\dagger(D^\mu H_i) -
M_0^2\left|H_{i+1}-\left(\frac{\Phi_{i+1}\phi_{i+1}^3}{(v_1/\sqrt{2})^3
v_2}\right)H_i\right|^2-V(H_i)\, , 
\label{eq:covderi}
\end{equation}
where $H_i$ is a doublet under $SU(2)_i$ and singlet for $SU(2)_{j\neq
i}$ with hypercharges $Y_i=1$ and $Y_{j\neq i}=0$. Following
\cite{Buras:2002ej} we parametrize its components as
\begin{equation}
H_i= 
\left[
 \begin{array}{c}
   i\chi_i^+\\
   \frac{1}{\sqrt{2}}(\psi_i-i\chi_i^3)
 \end{array}
\right]\qquad i=0,1,\ldots,N-1\, ,
\end{equation}
and the form of the potential it is chosen
$V(H_i)=-m^2H_i^\dagger H_i+\frac{\tilde{\lambda}}{2} (H_i^\dagger H_i)^2$.

The Yukawa sector, $\mathcal{L}_Y$, will be taken with the
Yukawa matrices independent of $i$
\begin{equation}
\label{eq:yukawas}
\mathcal{L}_Y=\sum_{i=0}^{N-1}\overline{Q_i} \widetilde{Y}_u H_i^c
U_i +\sum_{i=0}^{N-1}\overline{Q_i} \widetilde{Y}_d H_i
D_i+\hc \,,
\end{equation}
where $H_i^c\equiv i\tau^2H_i^\ast$ is the usual Higgs doublet
conjugate.

Finally, we choose to work in an arbitrary $R_\xi$-covariant gauge,
in the spirit of \cite{Herrero:1994tj}; this means that the $\pi$ fields 
will be maintained explicitly in our spectrum.\footnote{Alternatively, one may 
follow the approach of \cite{Hill:2000mu,Cheng:2001vd}, and remove the 
$\pi$ fields from the spectrum by resorting to a unitary-gauge type of 
gauge-fixing.}

Extracting the bilinear terms is straightforward but tedious.
The mass-eigenstate fields, before spontaneous symmetry breaking,
will be denoted by a ``tilde'' and are 
related to the gauge-eigenstate ones by 
\[
Q_{iL} = a_{ij}\widetilde{Q}_{jL}, \quad
U_{iR}=a_{ij}\widetilde{U}_{jR},\quad
Q_{iR} = b_{ij}\widetilde{Q}_{jR}, \quad 
U_{iL}=b_{ij}\widetilde{U}_{jL},
\]
\[
W_{\mu i}^a = a_{ij} \widetilde{W}_{\mu j}^a, \quad
\pi_i^a = b_{ij} \widetilde{W}_{5 j}^a, \quad
H_i = a_{ij} \widetilde{H}_j,
\]
where $a_{ij}$ and $b_{ij}$ are $N\times N$ and  $(N-1)\times (N-1)$ matrices, respectively,
given by 
\begin{equation}
\label{eq:defch1}
b_{ij}=\sqrt{\frac{2}{N}}\sin\left(ij\frac{\pi}{N}\right), \qquad\qquad
a_{ij}=\left\{ \begin{array}{rcl}j=0&\ & \sqrt{1/N}
\\ j \neq 0 & &
\sqrt{{2}/{N}}\cos\left(\frac{2i+1}{2}\frac{j\pi}{N}\right)\end{array}\right.\ok
\end{equation}
and the vector-like fields are defined as
$\widetilde{Q}_{i}=\widetilde{Q}_{iR}+\widetilde{Q}_{iL}$, and similarly
for $\widetilde{U}_i$.
The masses of the gauge bosons, to be denoted by 
$M_i$, 
are given by 
\begin{equation}
\label{eq:massdefinition}
M_i=2 \tilde{g}v_2 \sin\left(\frac{i \pi}{2N}\right)\,.
\end{equation}
In the limit  $N\to\infty$ one recovers the spectrum of the KK-modes 
of a continuous extra dimension, i.e. $M_i =  \tilde{g} v_2 i \pi/N$, provided that  
one identifies the length of the extra dimension as 
$\pi R \approx N/\tilde{g} v_2$. For a finite $N$ one can define 
$d\equiv 1/(\tilde{g}v_2)$ and $\pi R \equiv (N-1) d$, where $\pi R$ can be 
identified with the length of a discretized extra dimension with $N$ sites and $d$ as the 
lattice spacing. For $N=1$ we recover the SM. 
Note that the massless vector bosons
$\widetilde{W}_{\mu 0}^a$ and $\widetilde{B}_{\mu 0}$ are associated
to the SM model gauge bosons, i.e. 
they are the gauge bosons of the
unbroken diagonal group, which is identified with
the SM gauge group.
The continuum limit of the fermion masses is obtained by setting
$M_f=d^{-1}$ and their values coincide with those of the gauge bosons.  
In addition, the Higgs masses squared are given by
$M^2(\widetilde{H}_i)= M_i^2 - m^2 $.
Eventually $\widetilde{H}_0$ will break the gauge
symmetry spontaneously;  in fact it will be identified with the SM Higgs doublet,
therefore $\langle 0|\widetilde{H}_0|0\rangle=v/\sqrt{2}$ with
$v=246~GeV$. 

After symmetry breaking further diagonalizations 
are required in order to 
determine the final spectrum and mass-eigenstates of the theory.
One can verify that, at least for the
degrees of freedom we will be interested in, the spectrum of this
model coincides with that of a continuous universal extra dimension with
$m_n=n/R$ replaced by $M_n$. 
In particular, the mass $M(\widetilde{W}_{\mu i}^\pm)$
of the charged eigen-states $\widetilde{W}_{\mu i}^\pm$
is given by $M(\widetilde{W}_{\mu i}^\pm)=\sqrt{M_W^2+M_i^2}$.
The corresponding would-be Goldstone bosons $G^{\pm}_i$ 
and the  physical scalar $a^{\pm}_i$ are given by 
\begin{equation}
\label{eq:bosonsmixing}
G^{\pm}_i  =  \frac{M_i \widetilde{W}_{5i}^\pm+M_W
  \widetilde{\chi}^\pm_i}{\sqrt{M_i^2+M_W^2}} 
\stackrel{g\to 0}{\longrightarrow} \widetilde{W}_{5i}^\pm \,,\quad\quad
a^{\pm}_i  = \frac{-M_W \widetilde{W}_{5i}^\pm+M_i
  \widetilde{\chi}^\pm_i}{\sqrt{M_i^2+M_W^2}}  
\stackrel{g\to 0}{\longrightarrow} \widetilde{\chi}_{i}^\pm
\end{equation}
In the gaugeless limit, the Goldstone bosons and the physical scalars
can be directly identified with $\widetilde{W}_{5i}^\pm$ and
$\chi_{i}^\pm$, respectively.
Notice also that tree-level mixings between gauge-bosons and would-be Goldstone bosons
are removed through the choice of the $R_\xi$ gauge \cite{Buras:2002yj,Muck:2001yv}.

The determination of the final spectrum and the corresponding mass-eigenstates 
in the fermionic sector is slightly more involved; the details 
are presented in the  Appendix A.  
Note however  
 that for the rest of this paper we will only use 
the ``tilded'' fermion fields 
$\widetilde{Q}_{i}$ and $\widetilde{U}_{i}$
(i.e. the fields right before breaking the last 
$SU(2) \times U(1)_{Y}$). 
The fact that the $\widetilde{Q}_{i}$ and $\widetilde{U}_{i}$ are not 
eigenstates of the Lagrangian after the final symmetry breaking reflects
itself in the form of the tree-level propagators:

\begin{equation}
\label{eq:PropagatorTilde}
\left[
\begin{array}{cc}
\includegraphics[scale=0.6]{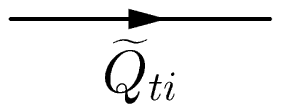}\hspace{3ex}
&
\includegraphics[scale=0.6]{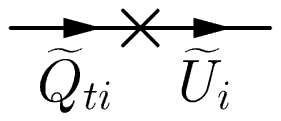}\hspace{3ex}\\
\includegraphics[scale=0.6]{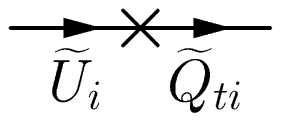}
&
\includegraphics[scale=0.6]{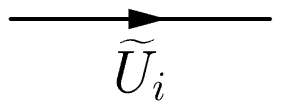}\hspace{3ex}
\end{array}
\right]
=
\left[
\begin{array}{cc}
\;i \frac{\slash{p}+M_i}{p^2-m_Q^2} &i \frac{m_t}{p^2-m_Q^2}\\
i \frac{m_t}{p^2-m_Q^2}&\;i \frac{\slash{p}-M_i}{p^2-m_Q^2}
\end{array}
\right]\ok
\end{equation}
where $m_Q^2=M_i^2+m_f^2$, 
and the ``cross'' denotes tree-level mixings, proportional 
to $m_f/M_i$ (see Appendix A). 

We end this section presenting the interaction terms 
relevant for our calculations; note in particular that  
we work in the limit of large top-quark mass.

From the Yukawa interaction, the couplings proportional to $m_t$ are 
\begin{equation}
\label{eq:yuklued}
\mathcal{L}_Y=\sum_{i=1}^{N-1} im_t\frac{\sqrt{2}}{v}
V_{tb}\overline{\widetilde{U}}_{tiR} \chi_i^+b_L+\hc \,.
\end{equation}

The couplings with the $Z$-boson
can be obtained from
$\mathcal{L}_Z = (g/2c_w) [J^{\mu}_{SM} + J^{\mu}_{F} + J^\mu_{\chi}] Z_\mu^{(0)}$, 
where $J^{\mu}_{SM}$ is the usual SM current,  and 
\begin{eqnarray}
\label{eq:sourcen}
J^{\mu}_{F}&=&\sum_{i=1}^{N-1}\left(1-\frac{4}{3}
s_w^2\right)\overline{\widetilde{Q}}_{ti} \gamma^\mu
\widetilde{Q}_{ti} -\frac{4}{3}
s_w^2\overline{\widetilde{U}}_{it}\gamma^\mu \widetilde{U}_{it} \,,\nonumber\\
J^{\mu}_{\chi} &=& \sum_{i=1}^{N-1} (-1 +2s_w^2)\widetilde{\chi}^{+}_i i\partial^\mu\widetilde{\chi}^{-}_i
+\hc
\end{eqnarray}

For the couplings with the photon, it easy to check that the
electromagnetic current can be written as 
$j_{em}^\mu =  j_{SM}^\mu+j_F^\mu+j_\chi^\mu$ \,,
where the new currents are
\begin{eqnarray}
\label{eq:emsource}
j^{\mu}_{F}&=&\sum_{i=1}^{N-1}\frac{2}{3}\overline{\widetilde{Q}}_{ti}
\gamma^\mu \widetilde{Q}_{ti} -\frac{1}{3}
\overline{\widetilde{U}}_{it}\gamma^\mu \widetilde{U}_{it}\,,\nonumber \\
j^{\mu}_{\chi} &=& \sum_{i=1}^{N-1} -\widetilde{\chi}^{+}_i
i\partial^\mu\widetilde{\chi}^{-}_i +\hc
\end{eqnarray}
\section{Bounds on the new physics}
\label{Bounds}
In this section we will set lower bounds on the mass of the lightest new 
particles of the model,  $M_1$, as defined in \Eq{eq:massdefinition},
which for simplicity is going to be denoted $M$ in what follows. 
As commented in the introduction 
in models with universal dimensions, continuous or latticized,  
due to the KK-number conservation
the only effects of new particles at low energies appear at 
the one-loop level. 
Since the new physics decouples when
$M\rightarrow\infty$,  the new contributions must scale as some 
inverse power of the new  physics scale, which must be compensated by
another scale from within the SM. The largest available scale in the SM is the 
top-quark mass. Thus, we expect large contributions in all observables 
that involve the top quark in loops. In particular, the new contributions 
to such observables will be suppressed only by factors 
$m^2_t/M^2$ with respect to the SM contributions.

In the SM there are several well-measured  
observables which are
very sensitive to the top-quark mass: the
decay rates $b \rightarrow s\; \gamma$ and $Z\to b\overline{b}$, the
$\rho$ parameter, and the rates of $B^0\rightleftharpoons
\overline{B^0}$.  
These observables will next be computed within the model we consider,  
with the expectation that they will turn out to be rather sensitive 
to the top-quark mass. 
\subsection{Radiative corrections to $b\to s \gamma$}
\label{subsec:bsg}
The experimental observable is the semi-inclusive decay $\mbox{B}(B\to
X_s\gamma)$. Using the heavy quark expansion it is found that, up to small
bound state corrections, this decay agrees with the parton model rates
for the underlying decays of the $b$ quark
\cite{Falk:1994dh,Neubert:1994ch}, $b\to s \gamma$. This flavor
violating transition is a very good place to look for new physics,
because in the SM it is forbidden at tree level due to gauge symmetry,
thus it can only proceed through radiative corrections. 
The transition can be parametrized by the following
effective Hamiltonian
\begin{equation}
\label{eq:effhamiltonian}
\mathcal{H}_{eff}=\frac{4G_F}{\sqrt{2}}V_{ts}^\ast V_{tb}\sum_{i=1}^8 C_i\mathcal{O}_i \,,
\end{equation}
where $\mathcal{O}_7$ is the operator that drives the transition
$b\to s \gamma$, 
\begin{equation}
\mathcal{O}_7= \frac{e}{(4\pi)^2}m_b \overline{s}\sigma^{\mu\nu} P_R b F_{\mu\nu}\,,
\end{equation}
and $C_7$ is a coefficient to be computed in the specific model. 
In the SM, and at the scale of the $W$-boson mass, 
$C_7^{SM}(M_W)=-1/2~A(m_t^2/M_W^2)$, where $x_t=m_t^2/M_W^2$, and 
\begin{equation}
\label{eq:defa}
A(x)=x\left[ \frac{\frac{2}{3}x^2 + \frac{5}{12}x
-\frac{7}{12}}{(x-1)^3} - \frac{\left(\frac{3}{2}x^2-x\right)\ln
x}{(x-1)^4} \right]\,.
\end{equation}
The other operators included in (\ref{eq:effhamiltonian}) do not contribute 
directly to $b\rightarrow s \gamma$; however, QCD radiative corrections mix
all the operators, and  $\mathcal{O}_7$ end up receiving contributions from 
the other operators as well. 
These contributions can be conveniently calculated, resummed
by using the renormalization group, and encapsulated in the evolution of the
Wilson coefficients $C_i(\mu)$ from $M_W$ to $m_b$.
It turns out that the corrections are numerically important
\cite{Agashe:2001xt,Buchalla:1996vs}:
\begin{equation}
\label{eq:running}
C_7(m_b)\approx 0.698\;C_7(M_W)-0.156\;C_2(M_W)+0.086\;C_8(M_W) \,,
\end{equation}
where $C_2$ and $C_8$ are the coefficients of the operators
\begin{subequations}
\begin{eqnarray}
\mathcal{O}_2 & = &[\overline{c}_{L\alpha} \gamma_\mu
b_{L\alpha}][\overline{s}_{L\beta} \gamma^\mu c_{L\beta}]\,,\\
\mathcal{O}_8 & = & \frac{g_s}{(4\pi)^2} m_b
\overline{s}_{L\alpha}\sigma^{\mu\nu} T^a_{\alpha\beta} b_{R\beta}
G^a_{\mu\nu}\,,
\end{eqnarray}
\end{subequations}
and $\alpha$, $\beta$ are color indices. 
In the case of the SM, the contribution of $\mathcal{O}_8$ is
negligible because it is generated at one loop, 
$C^{SM}_8(M_W)=-0.097$ \cite{Buchalla:1996vs}, but that 
of $\mathcal{O}_2$, $C^{SM}_2(M_W)=1$, is important because it 
appears at tree level.
Equation (\ref{eq:running}) is valid for any theory, as long as one assumes
that there is no new physics below $M_W$. Then, the running from $M_W$ to
$m_b$ is the standard QCD running, and all the new physics is included in the
boundary conditions for the Wilson coefficients at the scale $M_W$.

\begin{figure}
\begin{displaymath}
\begin{array}{c}
  \begin{array}{ccc}
     \includegraphics[scale=\currentscalevalue]{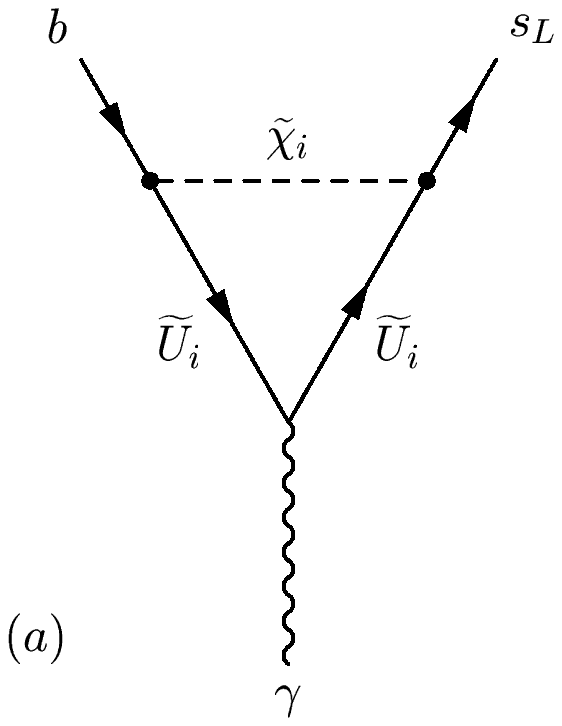}   &
     \includegraphics[scale=\currentscalevalue]{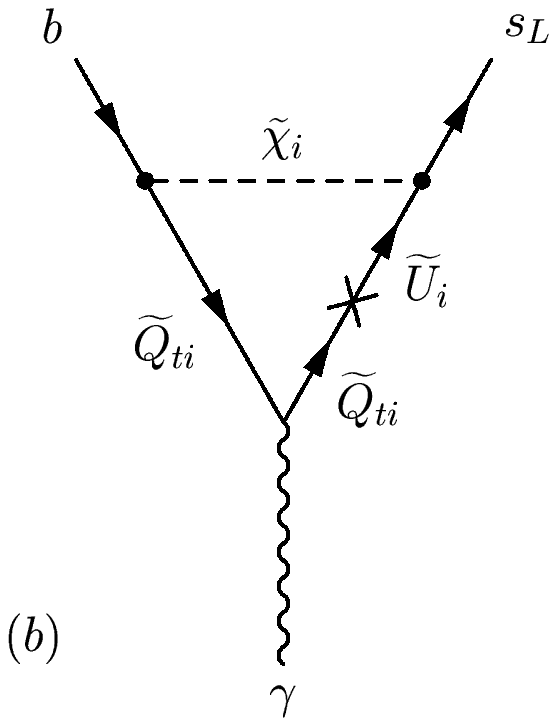}  &
     \includegraphics[scale=\currentscalevalue]{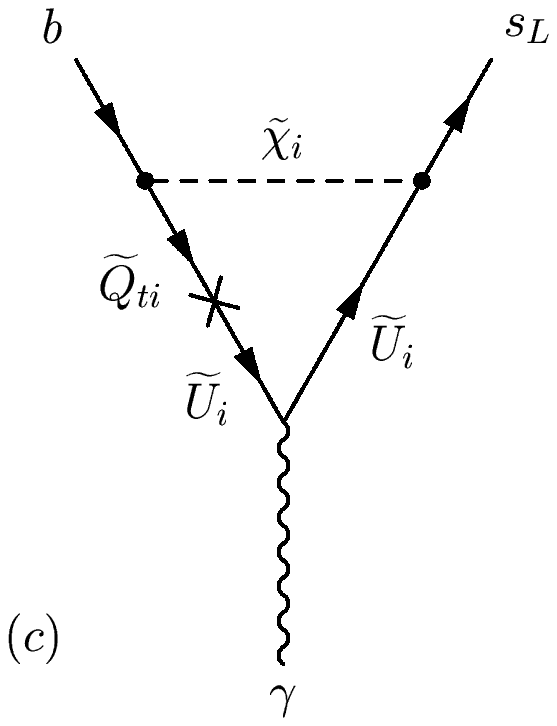}
  \end{array}
  \\
  \begin{array}{cc}
    \includegraphics[scale=\currentscalevalue]{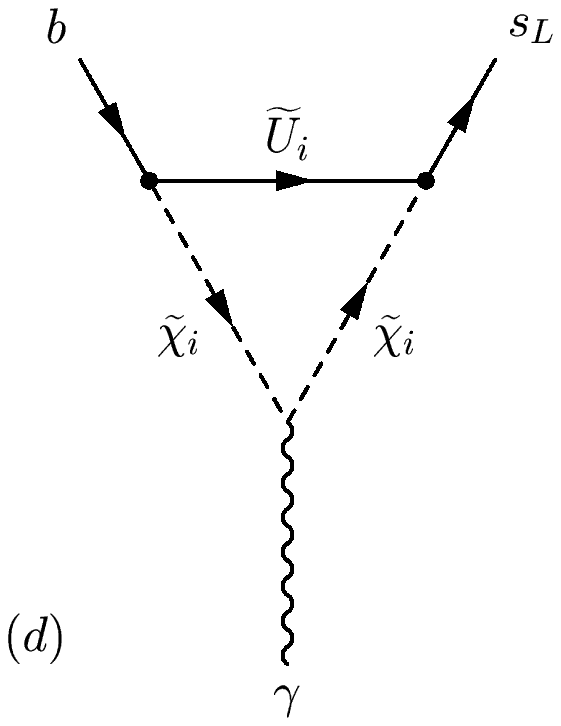}&
    \includegraphics[scale=\currentscalevalue]{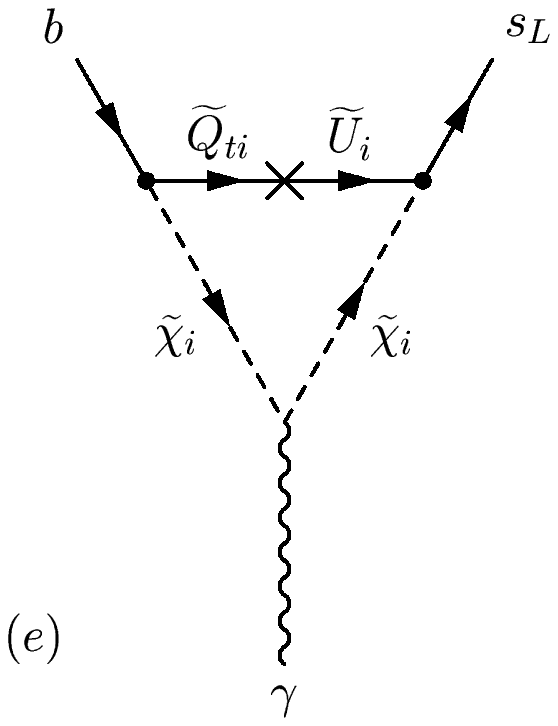}
  \end{array}
\end{array}
\end{displaymath}
\caption{Diagrams that contribute to $\mathcal{O}_7$.}
\label{fig:bsginLUED}
\end{figure}

In the model we consider, the transition  proceeds through the
same effective Hamiltonian, but the coefficient 
$C_7$ is modified by the diagrams of
\Fig{fig:bsginLUED}.
There are also diagrams in which the $\widetilde{\chi}_i$ are replaced
by $\widetilde{W}_{\mu i}$ and by the non physical degrees of freedom
$\widetilde{W}_{5i}$ but since the couplings of these are reduced by a
factor $(M_W/m_t)^2\approx~0.22$ we will ignore them and work at this
level of precision.

The contribution of the i-th mode to the $C_7$ coefficient
can be written in the form \cite{Agashe:2001xt}
\begin{equation}
\label{eq:c7inLUED}
C_{7\; i}(M_W)= \frac{r_i}{1+r_i} \left[ B\left(1+r_i\right) 
- \frac{1}{6} A\left(1+r_i\right) \right]~,
\end{equation}
where $r_i\equiv  m_t^2/M_i^2$, and $B(x)$ is given by
\begin{equation}
\label{eq:defb}
B(x)= \frac{x}{2} \left[ \frac{\frac{5}{6} x - \frac{1}{2}}{(x-1)^2} -
\frac{\left( x - \frac{2}{3}\right)\log x}{(x-1)^3} \right]~.
\end{equation}
An expansion
of $C_{7\;i}$ reveals that it does not contain logarithms of the
two different mass scales $M_i$ and $m_t$,
\begin{equation}
\label{eq:freelogs}
  C_{7\;i}(M_W)=\frac{23}{144} r_i -
  \frac{13}{120}r_i^2 + \mathcal{O}(r_i^3)~,
\end{equation}
a feature that can be understood from an effective field theory point
of view. Specifically, 
when the heavy degrees of freedom are integrated out, the
tree-level effective Lagrangian is exactly the SM Lagrangian; 
there are no additional tree-level operators suppressed 
by powers of $M_i^{-1}$. It
is well known that the dominant logarithms of the two different scales
can be recovered from the running of the
operators in the low energy effective Lagrangian induced by the
presence of the additional operators. Thus, since in our case there are no
additional tree-level operators, no logarithms can appear in \Eq{eq:freelogs}. 
As a matter of fact this is also an inherited property from UED, where 
no such logarithms appear either, for the same reason.
Finally, all contributions must be put together,
\begin{equation}
\label{eq:c7total}
C_{7}(M_W)=C_7^{SM}(M_W)+\sum_{i=1}^{N-1}
C_{7\;i}(M_W)\,,
\end{equation}
where we have neglected the running between $m_t$ and $M_W$, i.e.
$C_{7\;i}(m_t)\approx~C_{7\;i}(M_W)$.
 
Since the model we consider does not generate additional contributions 
at tree level, we basically obtain the SM result for $C_2$.
On the other hand, the $C_8$ coefficient can get corrections, 
which will be comparable to those of the SM; however, since the latter 
are negligible, so are the former.

For comparing the predictions for this process with the experimental
result, it is convenient to use the ratio
$\widetilde{\Gamma}=\Gamma(b\to
s\gamma)/\Gamma(b\to cl\nu)$, 
which depends much less on $m_b$, and, therefore, presents a smaller
uncertainty \cite{Kagan:1998ym}: 10\% for the theoretical value in the SM, 
while the uncertainty in the the experimental determination is about
15\% (both at $1\sigma$), and central values agree quite well with the SM
calculations. In fact, current determinations only allow for new 
physics contributions which are about 36\% of the SM value (at 95\%~CL) 
\cite{Agashe:2001xt}, i.e. 
$|\widetilde{\Gamma}^{total}/\widetilde{\Gamma}^{SM}-1|\leq 0.36$. 
Since the process $b\to c l \nu$ is only modified at one loop 
by the new physics the previous equation can be translated into 
the more useful one
\begin{equation}
\left| \frac{|C_7^{total}(m_b)|^2}{|C_7^{SM}(m_b)|^2} - 1
\right|<0.36 \qquad 95\%\;\mbox{CL}\,.
\end{equation}
The $C_7$ coefficients at the scale $m_b$ are obtained from
\Eq{eq:running}. The final bounds that one can set from this process are 
shown in \Fig{fig:results}.

\subsection{The $Z\to b\overline{b}$ process.}  
Radiative corrections coming from new
physics affect the branching ratio $R_b=\Gamma_b/\Gamma_h$, where
$\Gamma_b=\Gamma (Z\to b \overline{b})$ and $\Gamma_h=\Gamma (Z\to
\mbox{hadrons})$ and also the left-right asymmetry $A_b$. Both can be
treated uniformly by expressing them as a modification to the tree-level 
couplings $g_{L(R)}$ (it is understood that we refer only to the
couplings of the b-quark) defined as
\begin{equation}
\label{eq:definitions}
\frac{g}{c_W}\overline{b} \gamma^\mu(g_L P_L +g_R P_R)b Z_\mu \,,
\end{equation}
$Z$ and $b$'s are SM fields, $P_{L(R)}$ are the chirality projectors, and
\begin{subequations}
\begin{eqnarray}
g_L& = & -\frac{1}{2}+\frac{1}{3}s_W^2+\delta g_L^{SM}+\delta g_L^{NP}\,,\\
g_R & = &\frac{1}{3}s_W^2 + \delta g_R^{SM}+\delta g_R^{NP}\,,
\end{eqnarray}
\end{subequations}
where we have separated radiative corrections coming from SM
contributions and from new physics (NP). It turns out that, both
within the SM as well as in most of its extensions, only $g_L$
receives corrections proportional to $m^2_t$ at the one-loop level, due
to the difference in the couplings between the two chiralities. In
particular, a shift $\delta g_L^{NP}$ in the value of $g_L$ due to new
physics translates into a shift in $R_b$ given by
\begin{equation}
\label{eq:deltarb}
\delta R_b=2R_b(1-R_b)\frac{g_L}{g_L^2+g_R^2}\delta g_L^{NP}\,,
\end{equation}
and to a shift in the left-right asymmetry $A_b$ given by
\begin{equation}
\label{eq:deltaab}
\delta A_b=\frac{4g_R^2g_L}{(g_L^2+g_R^2)^2}\delta g_L^{NP}\,.
\end{equation}
These equations, when compared with experimental data, will provide
bounds on the new physics. A possible way of parametrize $\delta
g_L^{NP}$ is defining the function $F(a)$ through the relation
\begin{equation}
\delta g_L^{NP}=\delta
g_L^{SM}F(a)=\frac{\sqrt{2}G_Fm_t^2}{(4\pi)^2}F(a)\,,
\end{equation} 
where $a=\pi R m_t$ and $G_F$ is the Fermi constant. 

\begin{figure}[ht]
\begin{displaymath}
\begin{array}{c}
  \begin{array}{ccc}
    \includegraphics[scale=\currentscalevalue]{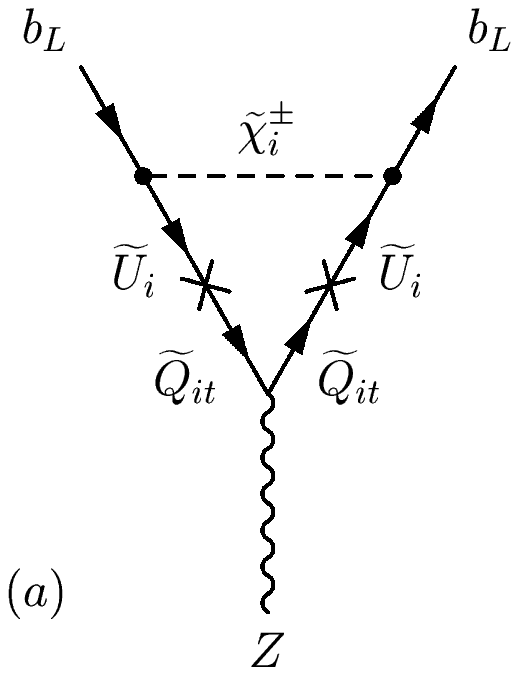}&
    \includegraphics[scale=\currentscalevalue]{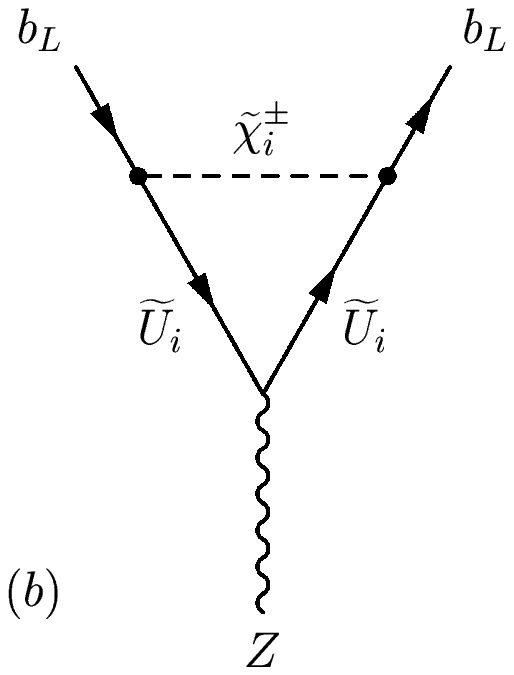}&
    \includegraphics[scale=\currentscalevalue]{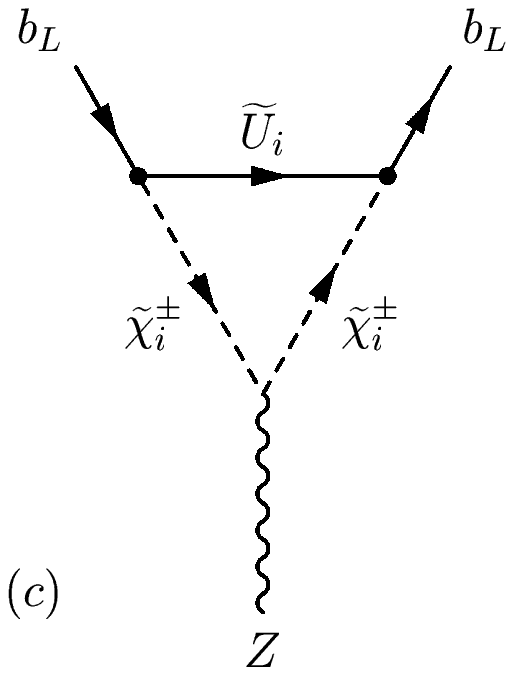}
  \end{array}\\
  \begin{array}{cc}
    \includegraphics[scale=\currentscalevalue]{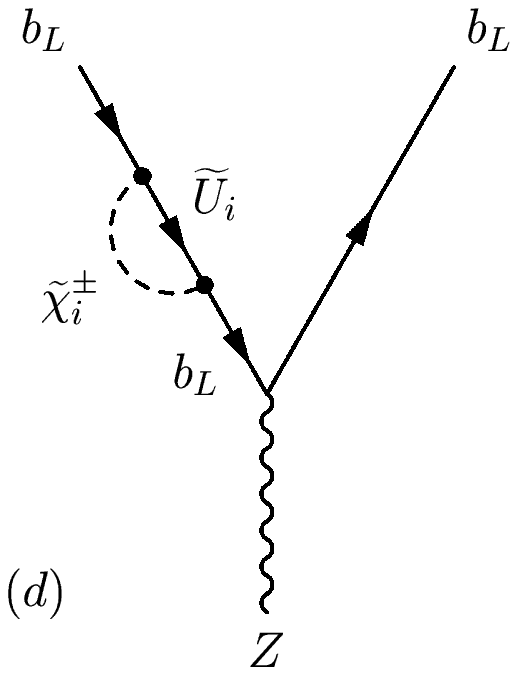}&
    \includegraphics[scale=\currentscalevalue]{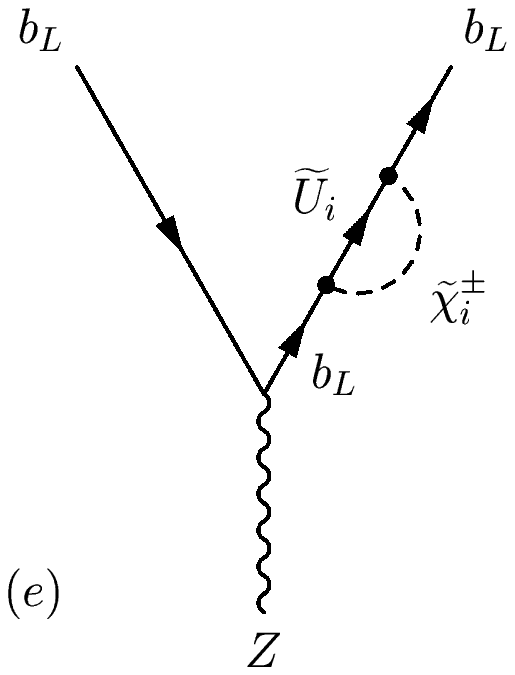}
  \end{array}
\end{array}
\end{displaymath}
\caption{Diagrams contributing to $Z\to b\overline b$.}
\label{fig:Zbb}
\end{figure}

In the model we consider, the new contributions stem from the set
of diagrams displayed in \Fig{fig:Zbb}.
 Following Ref.~\cite{Oliver:2002up} we parametrize the
different contributions as
\begin{equation}
i\mathcal{M}_i = i \frac{g}{c_w}
\frac{\sqrt{2}G_Fm_t^2}{(4\pi)^2}f(r_i) \overline{u}^\prime\gamma^\mu
P_L u  \epsilon_\mu \,,
\end{equation}
where $u$ and $u^\prime$ are the spinors of the $b$ quarks and
$\epsilon_\mu$ stands for the polarization vector of the $Z$
boson. Although each of the contributions is divergent, the sum is finite; 
the divergences cancel, and so do all terms proportional to
$s_w^2$. Thus, finally, the only term which survives is the term from
diagram~\Fig{fig:Zbb}a not proportional to $s_w^2$, yielding the following
contribution
\begin{equation}
\label{eq:deltagl}
\delta g_{Li} =\frac{\sqrt{2}G_Fm_t^2}{(4\pi)^2}
\left[\frac{r_i-\log(1+r_i)}{r_i}\right]\,.
\end{equation}
A way to understand this result is by resorting the so called gauge-less limit
\cite{Barbieri:1992nz,Barbieri:1993dq}, as was done in
Ref.~\cite{Oliver:2002up} in the context of the UED model. 
Notice also here the absence of logarithms in \Eq{eq:deltagl}
when $r_i\to 0$.
 
The full contribution 
$\delta g_L^{NP} = \sum_{i=1}^{N-1}\delta g_{Li}$
expressed in terms of $F(a)$ can be written in the form
\begin{equation}
F(a)=\int_0^1dx\sum_{i=1}^{N-1}\frac{a^2 x}{4 (N-1)^2
\sin^2(i\pi/ 2N)+a^2 x}\,.
\end{equation}
This function captures the correction proportional to $m_t^2$, the
full one-loop result could be adapted from \cite{Buras:2002ej} by
replacing $m_n\to M_i$ as explained above. 

Now we will extract the experimental bound on $F(a)$ and translate it
into bounds on $M$. The maximum experimentally allowed value of $F(a)$
could be extracted from the modifications induced to $R_b$ and $A_b$
via $\delta g_L^{NP}$, see
Eqs.~(\ref{eq:deltarb}--\ref{eq:deltaab}). The SM prediction for the
left-right asymmetry $A_b^{SM}=0.9347\pm 0.0001$ and the measured
value $A_b^{exp}=0.921\pm 0.020$ give a looser bound than the one
coming from $R_b$, $R_b^{SM}=0.21569 \pm 0.00016$ and
$R_b^{exp}=0.21664 \pm 0.00068$ which turns out to be $F(a)-1=-0.24\pm
0.31$. Making a weak signal treatment \cite{Feldman:1998qc} is easy
to obtain the 95\% CL bound $F(a)-1<0.39$, from which the results
displayed in \Fig{fig:results} follow.

\comment{The results for UED and HG
\cite{Papavassiliou:2000pq} can be easily derived from \Eq{eq:faued}
and \Eq{eq:fahg}
\begin{equation}
R^{-1}_{UED}> 230\;\mbox{GeV} \qquad R^{-1}_{HG}> 1\;\mbox{TeV}  \qquad 95\%\;\mbox{CL}
\end{equation}
Observe that in the case of UED the bound is perfectly compatible with
the one obtained from the $\rho$ parameter. Taking into account the
$m_t$ proportional corrections has not improved the bound although the
one it gives is not as loose as previously believed
\cite{Appelquist:2000nn}.

In the case of the latticized scenarios the bound to the first excited
mode as function of $N$ is extracted from \Eq{eq:falhg} and
\Eq{eq:falued} and it the results are shown in figure \ref{fig:LHG} in
dashed line.
}

\subsection{$\rho$ parameter}
\label{sec:rhocalculation}

The $\rho$ parameter can be defined as the ratio of the relative
strength of neutral to charged current interactions at low
momentum transfer. 
In the SM, and at tree level ,it is predicted to be unity 
as a consequence of the custodial symmetry of the Higgs potential:
\begin{equation}
\rho\equiv\frac{G_{NC}(0)}{G_{CC}(0)}\approx \frac{M_W^2}{c_W^2 M_Z^2}=1~.
\end{equation}
However, since the SM contains couplings that violate this symmetry, the Yukawa
couplings and the U(1) coupling $g'$, radiative corrections 
modify $\rho$. At one loop the $\rho$ defined above receives 
corrections from vertex, box and gauge-boson self-energy diagrams; however
the dominant contributions, proportional to $m_t^2$,
come from the top-quark loops inside 
the gauge boson self-energies. 
Keeping only these contributions, one has
\begin{equation}
\label{eq:defrho}
\rho=1+\frac{\Sigma_W(0)}{M^2_W}-\frac{\Sigma_Z(0)}{M^2_Z}
\approx 1+\frac{1}{M_W^2}\bigg(\Sigma_1(0)-\Sigma_3(0)\bigg)
\approx 1+N_c \frac{\sqrt{2}G_F m_t^2}{(4 \pi)^2}.
\end{equation}
$\Sigma_W(0)$ and $\Sigma_Z(0)$ are co-factors of the $g^{\mu\nu}$
in the one-loop self-energies of the $W$ and $Z$ bosons,
evaluated at $q^2=0$, and 
$\Sigma_1(0)$ and $\Sigma_3(0)$ are the equivalent functions for the
$W_1$ and $W_3$ components of the SU(2) gauge bosons. 
In arriving at the above formula one uses the fact that 
the photon-$Z$ self-energy $\Sigma_{AZ}^{\mu\nu}$ is transverse, 
i.e.  $\Sigma_{AZ} (0)=0$; this last property holds  {\it only} 
for the subset of graphs containing fermion-loops, but is no longer true
when gauge-bosons are considered inside 
the loops of $\Sigma_{AZ}$~\cite{Degrassi:1993kn,Papavassiliou:1996hj}. 
Finally, $N_c$ is the number of colors. 

\begin{figure}
\begin{displaymath}
\begin{array}{c}
\begin{array}{cc}
\includegraphics[scale=\currentscalevalue]{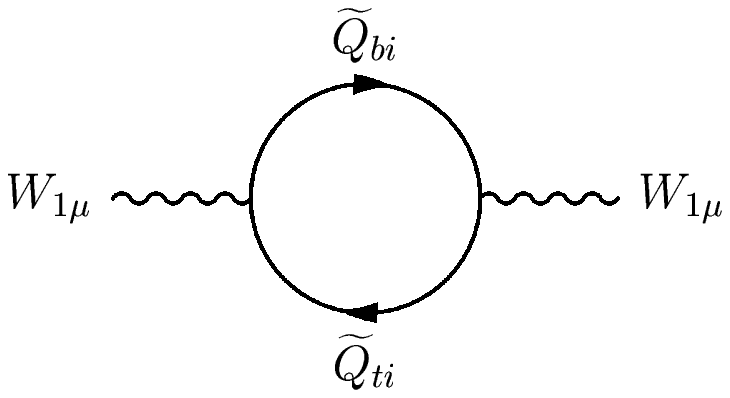}
&
\includegraphics[scale=\currentscalevalue]{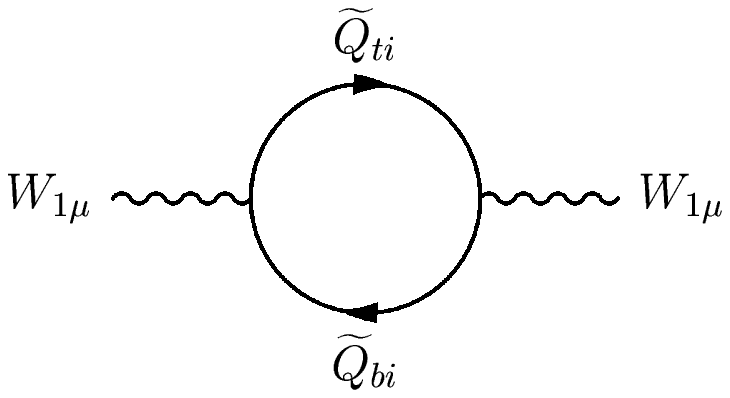}
\end{array}
\\
\includegraphics[scale=\currentscalevalue]{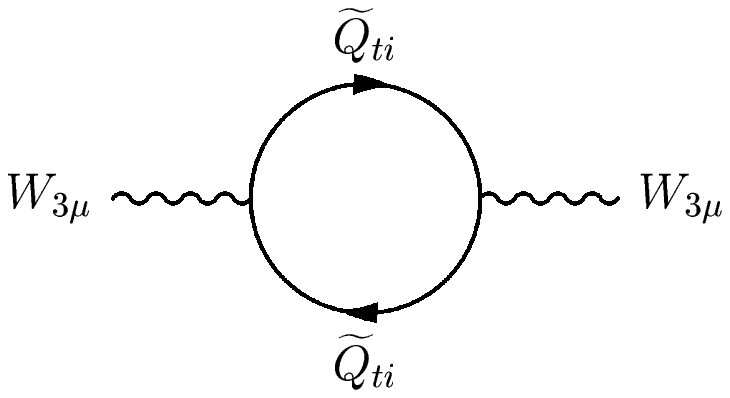}
\end{array}
\end{displaymath}
\caption{Diagrams modifying the $\rho$ parameter.}
\label{fig:rhodiags}
\end{figure}

Let us compute now the leading ($m_t^2$) corrections to $\rho$ in the model we consider.  
To that end we need the couplings of $W_1$ and $W_3$ components
(or alternatively the contributions of the new modes to the $J_1$ and $J_3$
currents),
\begin{equation}
\label{eq:ggaugea}
\mathcal{L}_{\rho}=\frac{g}{2}\sum_{i=1}^{N-1} W_\mu^1
\left[ \overline{\widetilde{Q}}_{it}\gamma^\mu
\widetilde{Q}_{ib}+\overline{\widetilde{Q}}_{ib}\gamma^\mu
\widetilde{Q}_{it}\right]+W_\mu^3 \left[
\overline{\widetilde{Q}}_{it}\gamma^\mu \widetilde{Q}_{it}\right]~,
\end{equation}
where we have already used the relation 
$g=\widetilde{g}/\sqrt{N}$~\cite{Cheng:2001vd}.
Thus, the couplings in
this basis are the same as in the SM, but the propagators of the fields 
are modified as described in the previous section.
The relevant diagrams are shown in \Fig{fig:rhodiags}.
Since only the $\widetilde{Q}$ fields couple to $W$ gauge bosons, we
use the component $Q$-$Q$ in the propagator of \Eq{eq:PropagatorTilde}.

The contribution to the $\rho$ parameter of each mode is finite and given by 
\begin{eqnarray}
\label{eq:finresl}
\Delta\rho_{i} &=&\frac{4}{g^2
v^2}\left[\Sigma_{1\;i}(0)-\Sigma_{3\;i}(0)\right]\\ &=& 2 N_c
\frac{\sqrt{2} G_F m_t^2}{(4\pi)^2}
\left[1-\frac{2}{r_i}+\frac{2}{r_i^2}\log(1+r_i)\right]\nonumber~.
\end{eqnarray}
The final result is found by summing the contributions of all the 
modes, $\Delta\rho=\Delta\rho^{SM}+\sum_{i=1}^{N-1}\Delta\rho_{i}$. In the
limit $N\rightarrow \infty$ we recover the UED result\cite{Appelquist:2002wb}.

\begin{figure*}
     \includegraphics[scale=1]{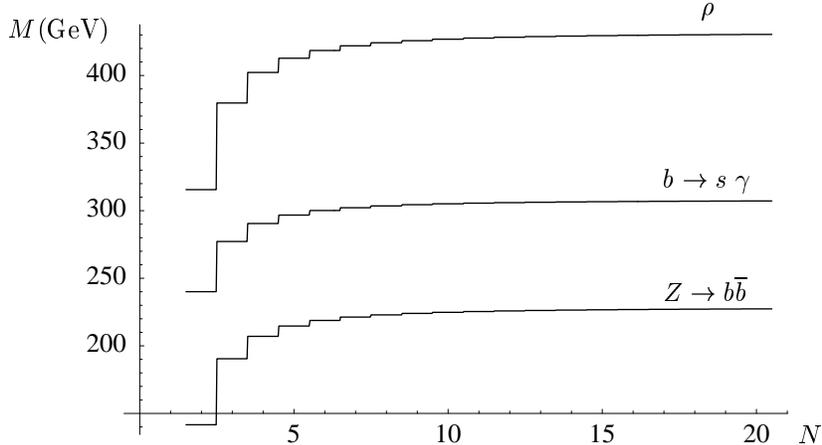}
\caption{Bounds on the mass scale of the new physics coming from
different precision observables as a function of the number of sites
$N$.}
\label{fig:results}
\end{figure*}

In order to discriminate between the corrections coming from the SM and
the ones coming exclusively from new physics we use the
$T$ parameter as defined in PDG \cite{Hagiwara:2002pw}. This
parameter contains only the corrections to $\rho$ coming
from new physics, $\alpha(M_Z) T \equiv \Delta \rho^{NP}$. 
$T$ is bounded to be $T<0.4$ at 95 \% CL. 
As in the UED case~\cite{Appelquist:2000nn}, 
this is the most restrictive observable.
In \Fig{fig:results} the resulting (95\% CL) bounds from the $\rho$ parameter
are displayed for different number of sites, $N$. 
In the limit of relatively large $N$ we find a limit of $430$~GeV for
the mass of the lightest new mode. This is in agreement with the latest 
results obtained in the UED model for a light Higgs 
boson.~\footnote{In Fig.~3 of \cite{Appelquist:2002wb} 
a limit of $500$~GeV at \%90 CL for $m_H=114$~GeV is quoted; this amounts
to a limit of about $400$~GeV at \%95 CL.}
If the number of sites is small, the 
bound can be reduced by a factor of about $10\%$--$25\%$, 
thus allowing new modes ranging between $320$--$380$~GeV.

\subsection{\kk\  mixing}
The model we consider has the same flavor structure as the SM, i.e. the
flavor violation is entirely governed by the CKM matrix. In models of this
type, both $\kk$ mixings and the
CP-violating parameter $\varepsilon_K$ can be parametrized
by a single function $S(x_t)$ \cite{Bergmann:2001pm}, with $x_t=m_t^2/M_W^2$. 
$S(x_t)$ is defined through
\begin{equation}
\mathcal{H}_{eff}^{\Delta B=2}= \frac{M_W^2 G_F^2 (V_{tb}
V_{td}^\ast)^2}{(4\pi)^2} S(x_t) [\overline{d}\gamma^\mu(1-\gamma_5)b]
[\overline{d}\gamma_\mu(1-\gamma_5)b]\,.
\end{equation}
In the SM
$S_{SM}(x_t)$ is dominated by the box diagrams with longitudinal $W$
exchanges and the top-quark running inside the loop
\begin{equation}
\label{eq:S}
S_{SM}(x_t)=\frac{x_t}{4} \left[ 1 + \frac{9}{1-x_t} - \frac{6}{(1-x_t)^2}
- \frac{6 x_t^2 \log(x_t)}{(1-x_t)^3}\right]\,.
\end{equation}

\begin{figure}
\begin{displaymath}
\begin{array}{cc}
\includegraphics[scale=\currentscalevalue]{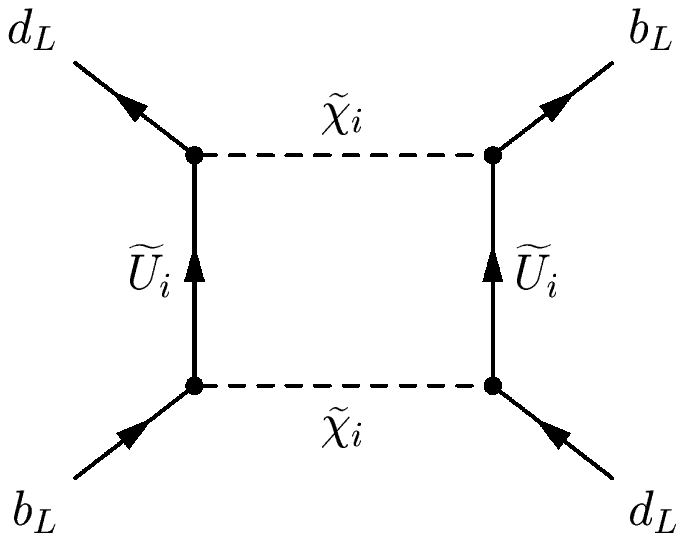}
&
\includegraphics[scale=\currentscalevalue]{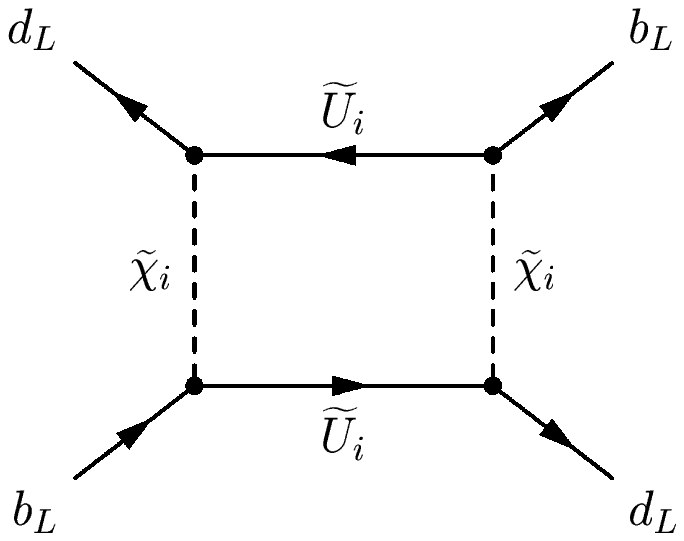}
\end{array}
\end{displaymath}
\caption{Diagrams for the dominant corrections to $S(x_t)$}
\label{fig:BBdiags}
\end{figure}

Using the running top-quark mass $m_t(m_t)=167\pm 5~\mbox{GeV}$, one obtains 
$S_{SM}(x_t)=2.39\pm 0.12$ \cite{Buras:2002yj}. If we split $S(x_t)$ in 
SM plus new physics  contributions, $S(x_t)=S_{SM}(x_t)+S_{NP}(x_t)$,
the latter can be encoded in a function defined as
$G(a)=S_{NP}/S_{SM}$. For our estimates we will use the limit of large top-quark mass. 
Then,  $S_{SM}(x_t)\approx x_t/4$, and
\begin{equation}
S_{NP}(x_t)=\frac{x_t}{4}(G(a)-1)\,. 
\end{equation}
The dominant contributions to this function
stem from the diagrams in \Fig{fig:BBdiags}.
Notice that only the
$\widetilde{U}_i$ fields are important because the couplings of
$\widetilde{Q}_i$ fields with the $\widetilde{\chi}_i$ are
proportional to the mass of the $b$ quark, and therefore provide
subdominant corrections. After the usual Fierz reordering and a bit of
combinatorics we obtain the result
\Eq{eq:sxtlued}. For comparison, we also present the results in the continuous
scenarios~\cite{Papavassiliou:2000pq,Buras:2002ej}:

\begin{eqnarray}
G(a)&=&1+2\int_0^1\sum_{i=1}^{N-1}\frac{a^2 x (1-x)\;dx}{4
(N-1)^2 \sin^2(i\pi/2N)+a^2 x}\,,\nonumber\\
\label{eq:sxt}
G_{UED}(a)&=&\int_0^1dx\;(1-x)\left[ a \sqrt{x} \coth(a \sqrt{x}
)+1\right] = 1+\frac{a^2}{18}- \frac{a^4}{540}+
\mathcal{O}(a^6) \,.
\label{eq:sxtlued}
\end{eqnarray}

The expression for $G(a)$ is obtained following arguments very
similar to the ones used in Ref.~\cite{Oliver:2002up}. The full 
one-loop calculation has been previously carried out in
Ref.~\cite{Buras:2002ej,Chakraverty:2002qk}, and reduces to \Eq{eq:sxt}
when only the dominant corrections, for large $m_t$, are retained;
the full one-loop result for $G(a)$ can be easily computed by
adapting the results in Ref.~\cite{Buras:2002ej}. 
Notice that the contributions in all cases are positive.

The last analyses of $S(x_t)$ \cite{Buras:2002yj} furnish a value which 
agrees well with the SM expectations
\begin{equation}
\label{eq:Sxtexp}
1.3\le S(x_t)\le 3.8  \qquad  95\% \,;
\end{equation}
the possible positive contributions have been lowered with respect to
previous determinations \cite{Bergmann:2001pm}, yielding better
bounds. 

Given the future improvements on the experimental determinations of
$\sin 2\beta$ by BaBar and BELLE, and in particular of the mass
splitting $\Delta M_s$ for the B sector in LHC and FNAL one may use
this observable to predict possible deviations from the SM
predictions. It turns out that, to an excellent accuracy
\cite{Buras:2002ej}, the aforementioned deviations in the case of $\Delta
M_s$ are governed by $G(a)$
\begin{equation}
G_{NP}(a)=\frac{(\Delta M_s)_{NP}}{(\Delta M_s)_{SM}}>1 \,.
\end{equation}
The larger values of $G(a)$ occur for small $N$, but they are
at most $G(a)\le 1.14$, which represents too small a deviation to be
discriminated experimentally. In fact, this observable is of the same order 
of magnitude as in the UED model~\cite{Buras:2002ej}.

The bounds one can set on the masses of
the new modes are below the $W$ mass, and are therefore irrelevant compared 
to previously discussed bounds.
The virtue of this results is
that the possible existence of extra latticized dimensions will not
pollute the extraction of the CKM matrix parameters from the future
improvements in the determination of the unitarity triangle.

\subsection{An alternative model with no new fermionic modes}

\begin{figure*}
     \includegraphics[scale=1]{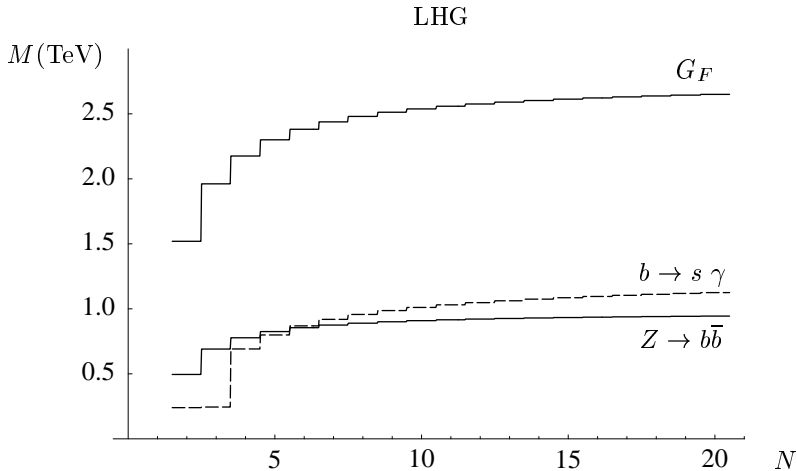}      
\caption{Bounds on the mass scale of the new physics in the
LHG models for different number of modes $N$.}
\label{fig:resultsLHG}
\end{figure*}

A popular scenario involving extra dimensions \cite{Pomarol:1998sd,Delgado:1999sv} 
is to assume  that only gauge bosons and Higgs scalars propagate in the
extra dimensions, while SM fermions are confined in the four dimensions. 
It is interesting, therefore, to study the  latticized version of such models, which
for brevity we will denote as LHG models.
The main difference compared to the universal models is that they  
generate new low energy physics already at tree level. In particular, the 
exchange of extra $W$ modes among SM fermions modifies the rate of muon decay,
and, therefore, the definition of $G_F$, while leaving unmodified, at tree
level, the well measured decay rates of the $Z$ boson. This mismatch can be
used to set stringent bounds on the masses of new particles in these scenarios
\footnote{The argument is the same in the continuous
scenarios\cite{Masip:1999mk, Nath:1999fs}.}.

The shift in $G_F$ produced by the modes $\widetilde{W}_{\mu\;i}$ is
\begin{equation}
\frac{G_F}{\sqrt{2}}=\frac{g^2}{8
  M_W^2}\left[1+\sum_{i=1}^{N-1}\frac{M_W^2}{M_i^2}2\cos^2\left(\frac{i\pi}{2 N}\right)\right]~,
\end{equation}
which reduces slightly the decay rate $\Gamma(Z \to \ell \overline
\ell)$
\begin{equation}
\frac{\Gamma^{LHG}}{\Gamma^{SM}}=1-\frac{(M_W\pi
R)^2}{2(N-1)^2}\sum_{i=1}^{N-1}\cot^2 \left(\frac{i
\pi}{2N}\right)\,.
\end{equation}
The maximum allowed deviation is very small \cite{Hagiwara:2002pw}, 
$|\Gamma^{NP}/\Gamma^{SM}-1|<0.0028$ at 95\%, 
and it provides very stringent bounds, shown in \Fig{fig:resultsLHG}, which are of the
order of a few TeV. For comparison we also display the limits we obtained
from $Z\rightarrow b\bar{b}$ and $b\rightarrow s \gamma$.


\section{Outlook and conclusions}
\label{Conclusions}
We have studied a five-dimensional extension of the SM in 
which the extra spatial dimension is latticized, and  
all SM fields propagate in it.
The model has the property that there are no tree-level effects below the
threshold of production of new particles. Therefore, 
to set a lower bound on the scale of the
new physics one should consider one-loop processes. We considered a number
of well-measured  observables, and which depend
strongly on the top-quark mass: 
the $\rho$ parameter, $b\to s \gamma$, $Z\to b\overline b$, and the
$B^0\rightleftharpoons\overline{B}^0$ mixing. 
The dominant
corrections, i.e. those proportional to the top-quark mass, have been computed,
and compared with the ones obtained when only the SM is
considered. 
It is found that the known bounds 
for the continuous version (UED) 
are rapidly reached when the extra dimension is
latticized by only about 10 to 20 (four dimensional) sites. However, when a
smaller number of sites is considered, the bounds on the scale of new physics is
lowered by roughly a factor of $10$\%--$25$\%, as can be seen in \Fig{fig:results}. Then,
the limits on new particles are about $320$--$380$~GeV. 
The bounds shown there correspond
to the mass of the lightest modes, defined in \Eq{eq:massdefinition}. 

We have also briefly discussed a latticized version of a model in which
fermions are confined to the four-dimensional subspace. In these models, 
there are contributions that modify the muon decay rate already at tree
level, allowing for much stronger bounds (of the order of $1$~TeV).

To summarize, we found that models with one universal latticized 
extra dimension provide new interesting physics which could be
well within the reach of the next generation of accelerators.

\begin{acknowledgments}
This work has been funded by the Spanish MCyT under the Grants
BFM2002-00568 and FPA2002-00612, and by the OCyT of the
``Generalitat Valenciana'' under the Grant GV01-94.
\end{acknowledgments}

\appendix
\section{\label{FME} The spectrum of the fermions}

After the spontaneous breaking of the remaining $SU(2)\times U(1)$, i.e. 
the usual SM symmetry group, the fermionic mass-eigenstates 
must be re-defined. In particular, the fermion masses will receive    
additional contributions from
the Yukawa piece $\mathcal{L}_Y$.
The Yukawa piece can be written in terms of the tilde fields as
\begin{equation}
\label{eq:yukpi}
\mathcal{L}_Y=\sum_{i=0}^{N-1}\overline{\widetilde{Q}}_i\frac{\widetilde{Y}_u}{\sqrt{N}}\widetilde{H}^c_0\widetilde{U}_i
+
\overline{\widetilde{Q}}_i\frac{\widetilde{Y}_d}{\sqrt{N}}\widetilde{H}_0\widetilde{D}_i+\hc\,,
\end{equation}
where we have concentrated on the terms containing the Higgs doublet
$\widetilde{H}_0$.
From the
first term in the sum of \Eq{eq:yukpi} is easy to convince oneself
that $Y_u\equiv\widetilde{Y}_u/\sqrt{N}$ is the SM Yukawa matrix. When
the Higgs doublet acquires a VEV one must diagonalize $Y_u$ using the
same field redefinitions as in SM, $\widetilde{Q}_{ui}\to U_u^\dagger
\widetilde{Q}_{ui}$, $\widetilde{U}_i\to V_u^\dagger
\widetilde{U}_i$. At the end the mass matrix for fermions will
be\footnote{We do not study explicitly the term containing $Y_d$
because it does not contain the mass of the top-quark, $m_t$, but its
treatment would be completely similar.}
\begin{equation}
\left(
  \begin{array}{cc}
    \overline{\widetilde{U}}_{if} & \overline{\widetilde{Q}}_{if}
  \end{array}
\right)
\left(
  \begin{array}{cc}
-M_i & m_f \\
m_f  & M_i
  \end{array}
\right)
\left(
  \begin{array}{c}
    \widetilde{U}_{if} \\\widetilde{Q}_{if}
  \end{array}
\right)~;
\end{equation}
$f$ is a flavor index, $f=u,c,t$, and $m_f$ are the masses of the SM up-type quarks.
The above mass matrix coincides
with the one obtained in the analogous model with one continuous extra dimension, after
making the substitution $M_n\to m_n$ \cite{Appelquist:2000nn}; 
$m_n=n/R$ is the mass of the n-th KK mode of the field in
the absence of Yukawa couplings. As a consequence, the mixing between
the $\widetilde{Q}$ and $\widetilde{U}$ is the same as in the aforementioned
continuous case. Denoting the new  mass eigenstates by ``primes'' we have that
\begin{equation}
\left(
  \begin{array}{c}
    \widetilde{U}_{if} \\
    \widetilde{Q}_{if}
  \end{array}
\right) = 
\left(
  \begin{array}{cc}
    -\gamma^5\cos\alpha_{if} & \sin\alpha_{if} \\
    \;\gamma^5\sin\alpha_{if} & \cos\alpha_{if}
  \end{array}
\right) 
\left(
  \begin{array}{c}
    U_{if}^\prime \\ Q_{if}^\prime
  \end{array}
\right) \ok
\end{equation}
where $\tan(2\alpha_{if})=m_f/M_i$. The masses are given by
$M(Q^\prime_{if}) = \sqrt{M_i^2+m_f^2} \equiv m_Q$. 
Notice that the zero-th modes have exactly the same masses as in SM,
all of them coming purely from the Yukawa piece \Eq{eq:yukpi} which,
as said, for the zero-th modes coincides exactly with the SM Yukawa
sector. In fact the same happens for the rest of the pieces in the
Lagrangian and one can safely identify the zero-th tilded fields,
$\widetilde{Q}_{0L}$, $\widetilde{U}_{0R}$ and $\widetilde{D}_{0R}$,
with the SM ones.


\end{document}

%% file: comandaments.tex
\newcommand{\rf}[1]{(\ref{#1})}

\newcommand{\hc}{\ensuremath{\mbox{h.c.}} }

\newcommand{\Slash}[1]{\ensuremath{#1\hspace{-0.65em}/\hspace{0.5ex}}}

\renewcommand{\slash}[1]{\ensuremath{#1\hspace{-1ex}/\hspace{0.5ex}}}

